\documentclass[manuscript]{aastex}

\slugcomment{January 14, 2013; Accepted for Publication in the Astronomical Journal}

\shorttitle{Massive Star Formation at the Periphery of W~39}
\shortauthors{Kerton et al.}

\begin{document}

\title{Massive Star Formation at the Periphery of the Evolved Giant
  \ion{H}{2} Region W~39 }

\author{C. R. Kerton}
\affil{Department of Physics \& Astronomy, Iowa State University,
  Ames, IA 50011, USA}
\email{kerton@iastate.edu}

\author{K. Arvidsson}
\affil{Astronomy Department, Adler Planetarium, 1300 S.\ Lake Shore Drive,
  Chicago, IL 60605, USA}
\email{karvidsson@adlerplanetarium.org}

\and

\author{M. J. Alexander} 
\affil{Department of Physics \& Astronomy, University of Wyoming, 1000
  E. University, Laramie, WY 82071, USA}
\email{malexan9@uwyo.edu}

\begin{abstract}
We present the first detailed study of the large, $\sim 30$ pc
diameter, inner-Galaxy \ion{H}{2} region W~39. Radio recombination
line observations combined with \ion{H}{1} absorption spectra and
Galactic rotation models show that the region lies at $V_\mathrm{LSR}
= +65.4\pm0.5$ km~s$^{-1}$ corresponding to a near kinematic distance
of 4.5$\pm0.2$ kpc. Analysis of radio continuum emission shows that the
\ion{H}{2} region is being powered by a cluster of OB stars with a
combined hydrogen-ionizing luminosity of $\log(Q) \geq 50$, and that there
are three compact \ion{H}{2} regions located on the periphery of W~39,
each with $\log(Q)\sim 48.5$ (single O7 - O9 V star equivalent). In
the infrared, W~39 has a hierarchical bubble morphology, and is a
likely site of sequential star formation involving massive
stars. Kinematic models of the expansion of W~39 yield timescales of
order Myr consistent with a scenario where the formation of the
smaller \ion{H}{2} regions has been triggered by the expansion of
W~39. Using {\it Spitzer} GLIMPSE and MIPSGAL data we show that
star-formation activity is not distributed uniformly around the
periphery of W~39 but is concentrated in two areas that include the
compact \ion{H}{2} regions as well as a number of intermediate-mass
Class I and Class II YSOs.

\end{abstract}


\section{Introduction} \label{sec:intro}

W~39 is a large \ion{H}{2} region in the inner Galaxy ($l=19\degr$,
$b=-0\fdg4$) first identified in the \citet{wes58} 1.4 GHz radio 
continuum survey and more recently cataloged as a large infrared ``bubble''
by both professional astronomers \citep[N24;][]{chu06} and citizen
scientists \citep[MWP1G18908-003146;][]{sim12}. It is a significant
high-mass star-forming region and a possible example of sequential or
triggered star formation based on the hierarchical bubble structure seen
in the infrared, but, owing to its highly obscured location in the
inner Galaxy, it has been poorly studied. Recent high-resolution
(sub-arcminute) surveys at radio and infrared wavelengths have
provided an opportunity to examine this region in detail for the first
time. This paper presents an analysis of the \ion{H}{2} region itself
along with an investigation of the influence of the \ion{H}{2} region
on star formation in its vicinity.

In the next section the key parameters of the various data sets used
in the study are summarized. The determination of the radial velocity
and distance to the \ion{H}{2} region is described in
\S\ref{sec:rvd}. An analysis of W~39 and three associated compact
\ion{H}{2} regions follows in \S\ref{sec:ghii}. In \S\ref{sec:yso} the
young stellar object (YSO) distribution surrounding W~39 is explored,
and our conclusions are presented in \S\ref{sec:conc}.

\section{Observations}\label{sec:obs}

\subsection{Radio Line \& Continuum} \label{sec:radio}

At cm radio wavelengths we use VLA Galactic Plane Survey (VGPS;
\citealt{sti06}) 21-cm line and continuum data. The VGPS maps have an
angular resolution of $\sim1$~arcmin, and the \ion{H}{1} line spectra
have a velocity  resolution of 1.56 km~s$^{-1}$. The maps include
single-dish short-spacing data, making them ideal for obtaining flux
density measurements of large angular scale ($\ga 30$~arcmin) objects
like W~39. 

To trace molecular gas we use $^{13}$CO (J$=$1$-$0) observations from
the Boston University/Five College Radio Astronomy Observatory
Galactic Ring Survey (BU/FCRAO GRS; \citealt{jac06}). These data have
an angular resolution of 46~arcsec and 0.21 km~s$^{-1}$ velocity
resolution. We also make use of 1.1 mm continuum data from the Bolocam
Galactic Plane Survey (BGPS; \citealt{ros10}).

\subsection{Infrared} \label{sec:ir}

At infrared wavelengths we primarily use data from the GLIMPSE
\citep{ben03} \emph{Spitzer} legacy survey. GLIMPSE provides 1.7 to
2.0~arcsec resolution images in four bands centered at 3.6, 4.5, 5.8,
and 8.0 $\mu$m along with a point source catalog that contains near-
and mid-infrared photometric data. In addition, we use the 6~arcsec
resolution images at 24~$\mu$m from the MIPSGAL \emph{Spitzer} legacy
survey \citep{car09}.  For near-infrared photometry of point sources
we use data from the Two-Micron All Sky Survey (2MASS;
\citealt{skr06}) and the United Kingdom Infrared Telescope (UKIRT)
Infrared Deep Sky Surveys-Galactic Plane Survey (UKIDSS-GPS;
\citealt{luc08}).

\section{Radial Velocity and Distance}\label{sec:rvd}

\subsection{Radial Velocity} \label{sec:rv}

\begin{figure}
\plotone{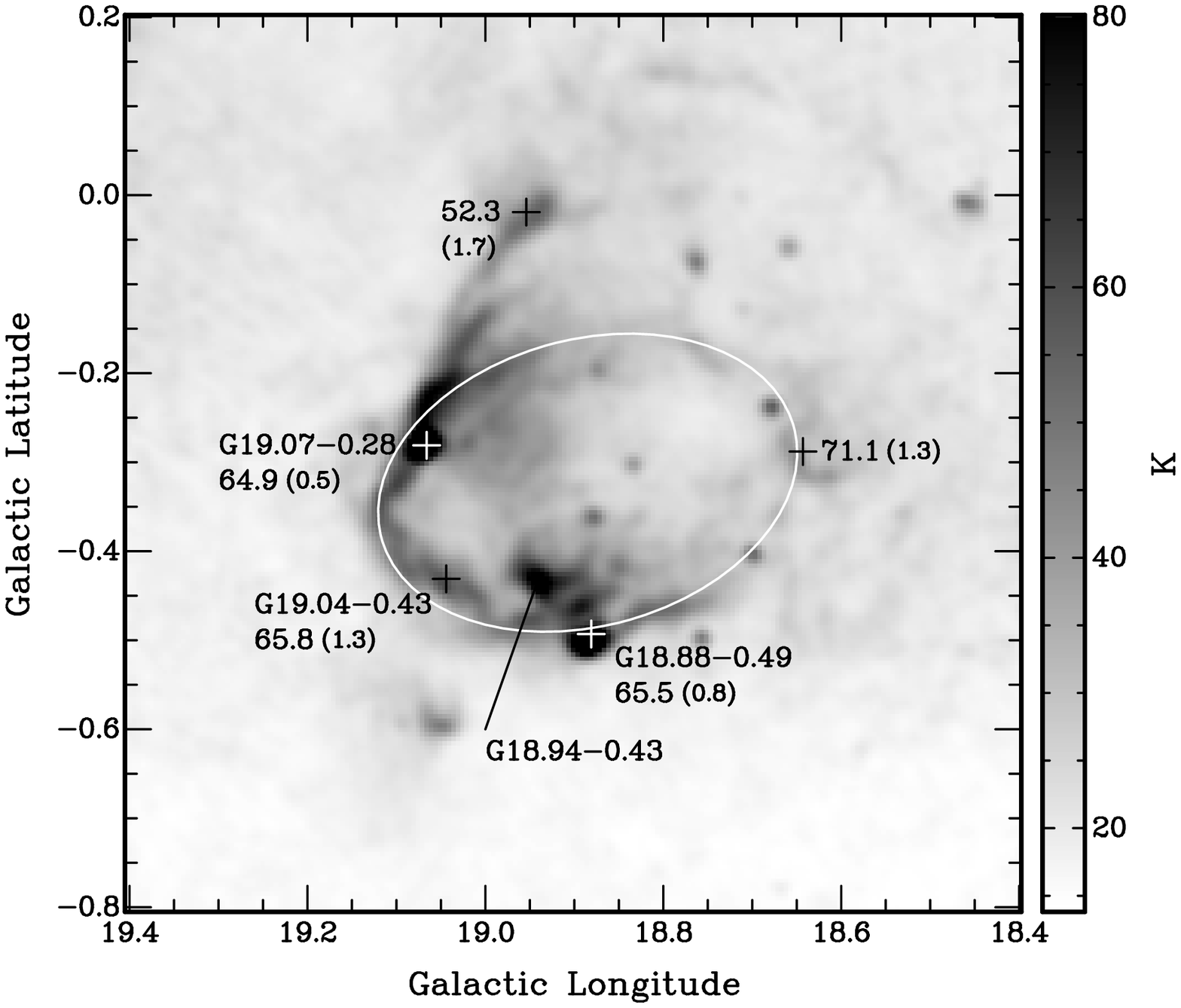}
\caption{VGPS 1420 MHz continuum image of the W~39 region. Crosses
  indicate the positions of RRL observations from \citet{loc89}. The
  peak velocity ($V_\mathrm{LSR}$, km~s$^{-1}$) of each RRL profile is
  also shown with the fit  uncertainty given  in parentheses. The position of
  the \ion{H}{2} region G18.94-0.43 is also shown.  The white ellipse
  approximately traces the position of the bubble structure best seen
  in the infrared (see Figure~\ref{fig:ir8_24}). \label{fig:w39cont}}
\end{figure}

In Figure~\ref{fig:w39cont} we show the VGPS 1420 MHz continuum map
of the area around W~39 along with the position and velocity of
cm-wavelength observations of the radio recombination lines (RRL)
H85$\alpha$, H87$\alpha$, and H88$\alpha$ \citep{loc89}. W~39 has a
bubble morphology best seen in the infrared. At radio wavelengths the
brightest emission ($\geq 50$~K) from W~39 is concentrated along a
U-shaped ridge opening to lower Galactic longitudes. Fainter emission
($\sim$ 30 -- 40~K) is found within the bubble and also tracing the
bubble edges at lower Galactic longitude. There is a local minimum in
the radio emission around $l=19\fdg025, b=-0\fdg38$ that is a likely
location for the exciting OB star(s). Three of the RRL observations
that lie on the strong ridge of emission have very similar velocities,
and we adopt the average of these values, $V_\mathrm{LSR}=65.4\pm0.5$
km~s$^{-1}$, as the radial velocity of W~39. The RRL velocity of
G18.95-0.02 is significantly different from the main ridge and, as we
discuss in \S\ref{sec:irmorph}, this region is likely not directly
associated with W~39.  

The nature of the RRL emission associated with the arc-shaped
structure G18.64-0.29 is more speculative. Both \citet{hel06} and
\citet{bro06} classify this object as a supernova remnant (SNR) based
on multi-frequency radio observations, and \ion{H}{1} absorption
studies by \citet{joh09} place the object at a distance of $\sim 5$
kpc. RRL emission is not expected from SNRs so it is likely that the
observed RRL emission, with a difference in velocity of only $\sim 5$
km~s$^{-1}$ from the main W~39 ridge, is originating from the edge of
the W~39 bubble.

\subsubsection{HI Absorption} \label{sec:hiabs}

\ion{H}{1} absorption spectra, obtained from the VGPS, were used to
determine if all three of the RRL sources are located at the same
near or far kinematic distance. A radio continuum source in the inner
Galaxy at the near distance can only show absorption up to its
$V_\mathrm{LSR}$ while a source at the far distance can show absorption at
velocities beyond $V_\mathrm{LSR}$ up to the tangent velocity, which at
$l\sim19\degr$ is 110 to 140 km~s$^{-1}$ depending on the kinematic
model used (see \S~\ref{sec:kindis}).

\begin{figure}
\plotone{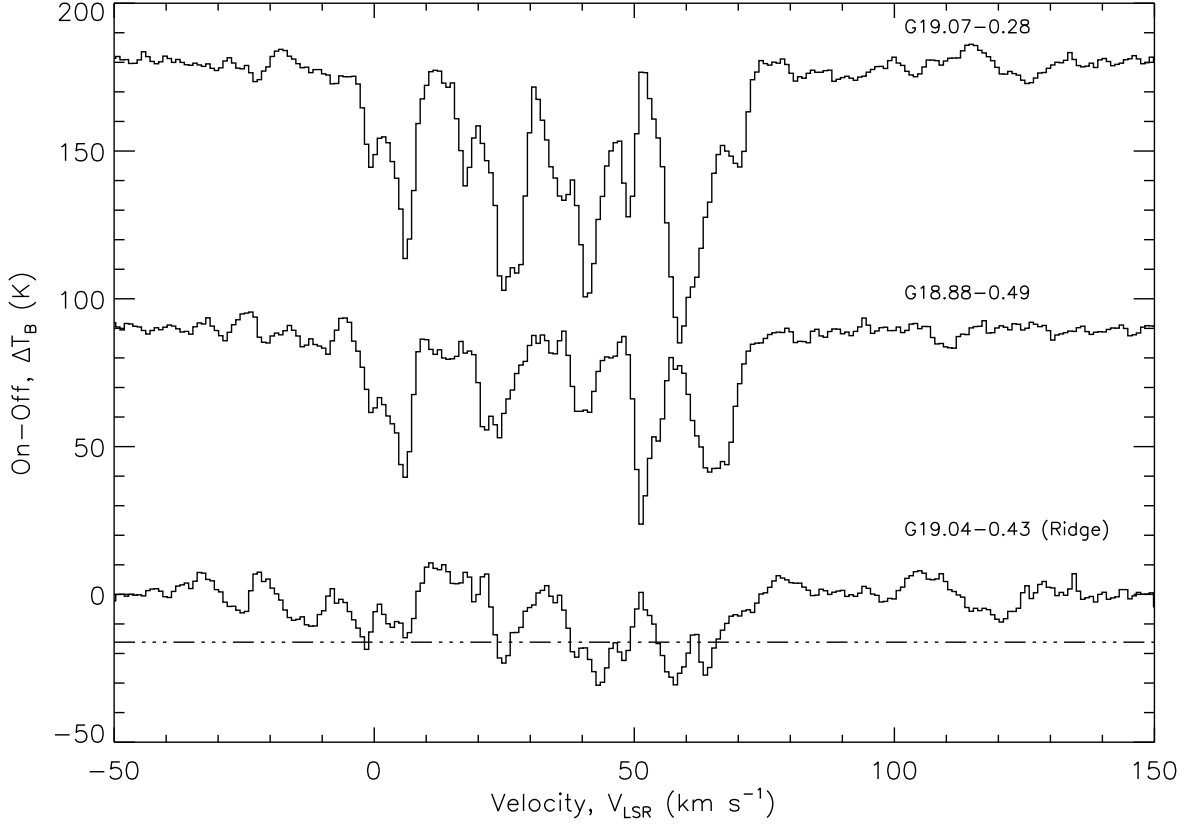}
\caption{\ion{H}{1} absorption spectra toward W~39. Spectra of two
  compact \ion{H}{2} regions (G19.07-0.28 and G18.88-0.49) and of a
  portion of the continuum ridge between the regions (G19.04-0.43)
  were obtained using data from the VGPS. The G18.88-0.49 and G19.07-0.28
  spectra have been offset by 90~K and 180~K for clarity, and the
  $3\sigma$ significance cutoff (dash-dot line) is shown for the
  G19.04 spectrum. At $l\sim19\degr$ the model-dependent tangent velocity is 110 to
  140 km~s$^{-1}$. Note the lack of any significant absorption signal
  beyond $\sim 65$ km~s$^{-1}$ in all three cases showing that W~39 is
  located at the near kinematic distance. \label{fig:hispec}}
\end{figure}

For each of the RRL sources \ion{H}{1} ``ON'' spectra were obtained by
averaging \ion{H}{1} spectra where the 1420 MHz continuum brightness
temperature was above a threshold of 100~K for G19.07-0.28, 100~K for
G18.88-0.49, and 55~K for G19.04-0.43. ``OFF'' spectra were then
calculated by taking the average \ion{H}{1} spectrum of a nearby (1--2
arcmin offset) similarly sized region. The resulting ON-OFF absorption
spectra are shown in Figure~\ref{fig:hispec}.

The uncertainty in an absorption spectrum $\sigma_\mathrm{abs}$ can be
quantified by taking the maximum of either the standard deviation in
channels where no absorption is expected to occur (e.g., at negative
$V_\mathrm{LSR}$ in this case) or by constructing ``OFF-OFF'' spectra that
measure the intrinsic variability of \ion{H}{1} emission structure. We
found $\sigma_\mathrm{abs}\sim 5$~K in each case.  Looking at
Figure~\ref{fig:hispec} we see that the two brightest RRL sources are
clearly at the near kinematic distance because of the lack of
absorption at velocities $\gtrsim 65$~km~s$^{-1}$. The ridge RRL source has a
much poorer quality spectrum, but there is no evidence for absorption
beyond $\sim 65$~km~s$^{-1}$. We conclude that all of the emission
associated with the three RRL observations is at the near kinematic distance.

\subsubsection{IR Morphology} \label{sec:irmorph}

\begin{figure}
\plotone{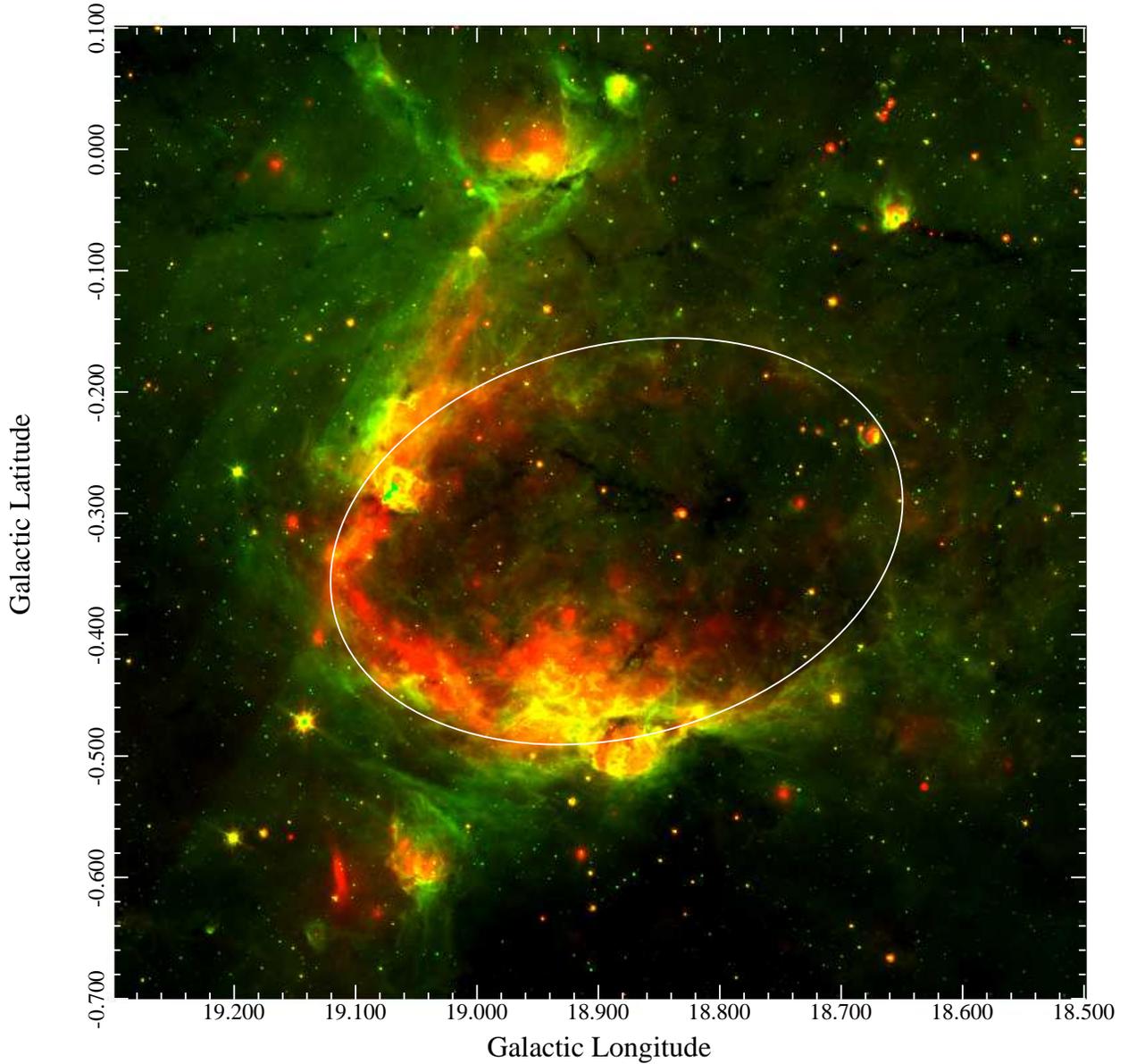}
\caption{GLIMPSE and MIPSGAL image of W~39. This image combines 8 and 24
  $\mu$m images from the GLIMPSE and MIPSGAL surveys (green and red
  respectively). The hierarchical bubble structure of W~39, consisting
  of the main bubble (white ellipse) and the two compact \ion{H}{2}
  regions G18.88-0.49 and G19.07-0.28, is best seen at 8.0 $\mu$m. As
  discussed in the text, the relative position of the 24 and 8 $\mu$m
  emission is consistent with W~39 being a single large \ion{H}{2}
  region. The white ellipse is the same as that shown in
  Figure~\ref{fig:w39cont}. \label{fig:ir8_24}}
\end{figure}

In Figure~\ref{fig:ir8_24} we show an 8 and 24 $\mu$m image of the
W~39 region using data from {\it Spitzer} GLIMPSE and MIPSGAL. The
infrared bubble structure, best seen at 8~$\mu$m and identified by
\citet{chu06} and \citet{sim12}, is indicated by an ellipse. As
expected for a thermal radio source W~39 is very bright in the
infrared. Of particular interest though is the layered structure of
the infrared emission; moving from the interior of the  U-shaped
structure to the exterior, we first see 24~$\mu$m emission followed by
a layer of 8~$\mu$m emission. This is exactly what would be expected
for a region being heated from one side by ionizing radiation. Within
the \ion{H}{2} region polycyclic aromatic hydrocarbons (PAH) are
strongly depleted allowing 24.0 $\mu$m emission from small dust grains
to dominate while in the photodissociation region (PDR) surrounding the
\ion{H}{2} region PAHs can survive resulting in strong 8~$\mu$m
emission \citep[e.g.,][]{tie93,gia94,ker08,arv11}. The continuity of this
layered infrared morphology is consistant with W~39 being a single
large \ion{H}{2} region. 

We note that this layered infrared morphology does not extend to
the region around $l=18\fdg95, b=-0\fdg01$. In Figure~\ref{fig:co} we
show GRS maps of $^{13}$CO emission between $0 \leq V_\mathrm{LSR} \leq
+130$~km~s$^{-1}$ integrated in 20~km~s$^{-1}$ increments
between 0 and +40 km~s$^{-1}$, 10~km~s$^{-1}$ increments
between 40 and +70 km~s$^{-1}$, and a single 60~km~s$^{-1}$
interval between +70 and +130 km~s$^{-1}$. The G18.95-0.01 region is
morphologically very similar to CO emission seen in the GRS at
$V_\mathrm{LSR} \sim +45$ km~s$^{-1}$ consistent with the lower RRL velocity
observed by \citet{loc89}. Based on this we do not consider this
region to be part of W~39. The CO maps also show that most of the
molecular gas along the line of sight to W~39 is found between +60 and
+70 km~s$^{-1}$.

\subsection{Kinematic Distance} \label{sec:kindis}

The (near) kinematic distance to W~39 was calculated using
$V_\mathrm{LSR}=65.4\pm0.5$ km~s$^{-1}$ and  three different Galactic
rotation models: 1) the R$_\sun$= 8~kpc hydrodynamical model of
\citet{poh08}, 2) the R$_\sun$=8.5~kpc, V$_\sun$=220 km~s$^{-1}$ model
of \citet{cle85}, and 3) a version of the \citet{cle85} model scaled
to R$_\sun$=8.0~kpc. The resulting distances are 4.5 kpc, 4.7 kpc, and
4.4 kpc respectively. In each case the error in the calculated
distance caused by the uncertainty in $V_\mathrm{LSR}$ is only $\sim0.04$
kpc, much less than the systematic uncertainty associated with the
choice of rotation model. We adopt the average (standard deviation) of
these values as the distance (uncertainty) to W~39:
$d=4.5\pm0.2$~kpc. At this distance, W~39 is probably part of the
Scutum-Centaurus Arm of our Galaxy \citep{chu09,rus03}.

\section{A Giant HII Region}\label{sec:ghii}

\subsection{Exciting Stars \& Ionized Mass}

The flux density at 1420 MHz of W~39, $F_{1420}=40\pm0.6$~Jy, was
measured from VGPS data using {\sc imview} \citep{hig97} to integrate
the emission within a user-defined area roughly following the
$T_B=30$~K contour, but omitting emission at $b>-0\fdg1$ and
subtracting the contribution from compact and point sources found
within the contour. The main contribution to the uncertainty comes
from the definition of the background level which varies depending on
the number and location of the points used to define it and the order
of the two-dimensional polynomial fit used.

The observed radio continuum flux density ($F_\nu$) of an \ion{H}{2}
region at a distance ($d$) can be related to the
ionizing radiation luminosity ($Q$) of the star(s) located within the
\ion{H}{2} region by:
\begin{equation} \label{eqn:logq}
\log\left(fQ\right) = 40.95 + \log\left(\frac{F_\nu}{\mathrm{Jy}}\right) + 2\log\left(\frac{d}{\mathrm{pc}}\right) ,
\end{equation}
where $f$ is the fraction of the ionizing radiation from the star(s)
that is intercepted by the \ion{H}{2} region (e.g., $f=1$ for an
ionization bounded region, $f\sim0.5$ for a ``blister'' \ion{H}{2}
region), and we used $\nu = 1.4$~GHz and $T_e = 7500$~K to calculate
the constant \citep{ker99, mat76}. As pointed out by \citet{rub68}
Equation~\ref{eqn:logq} does not depend on the details of the density
structure of the \ion{H}{2} region.

For $d=4500$~pc and $F_\nu=40$~Jy we find $\log\left(fQ\right) = 49.86$,
equivalent to the output of two O3~V stars (or 10 O7~V stars) using
the calibration of \citet{cro05}. Varying distance and $F_\nu$ within
the uncertainties, and allowing $T_e$ to range between 7500~K and
10,000~K, we find the full range of $\log\left(fQ\right) = 49.75 -
49.90$.  Given the morphology of the \ion{H}{2} region it is very
likely that $f<1$  and W~39 is a giant \ion{H}{2} region
\citep[GHIIR;][]{con04} being powered by numerous OB stars.

The formulas given in \citet{mat76}, and derived in detail in
\citet{mez67}, can be used to estimate the average electron density of
the \ion{H}{2} region and the total mass of ionized gas given $F_\nu$,
$d$, $T_e$, and the angular size of the region. An appropriate size
scale for W~39 is given by the separation between the two sides of the
\ion{H}{2} region, $\sim 0\fdg4$, or 30 pc at 4.5 kpc. This is
slightly smaller than the value of $0\fdg6$ reported in \citet{wes58},
but that value included emission at higher Galactic latitude that we
do not associate with the \ion{H}{2} region. We find values of $n_e
\sim 20$~cm$^{-3}$ and $M\sim10^{4}$~M$_\sun$. Unlike the
$\log\left(fQ\right)$ calculation these values do depend strongly on
the details of the \ion{H}{2} region's density structure and should
only be considered order-of-magnitude estimates.

\subsection{G19.07-0.28 = IRAS 18239-1228}

The radio source G19.07-0.28 is the brightest of the compact radio
sources found on the periphery of W~39 with
$F_\nu=1.77\pm0.08$~Jy. Using the same formulas and calibrations from
the previous subsection we find $\log\left(fQ\right)=48.50$ equivalent to a
single O8~V star. The ionized gas mass is $M\sim 50$~M$_\sun$ and the
average electron density is $n_e\sim160$~cm$^{-3}$. This radio source
is also the infrared source IRAS 18239-1228 and has four well-defined
IRAS flux density measurements. Using the technique of \citet{eme88}
we calculate the far-infrared luminosity $\log(L_{\rm IR}/L_\sun) =
5.2\pm0.1$ which is the luminosity of an O7 - O8~V star. 

\begin{figure}
\plotone{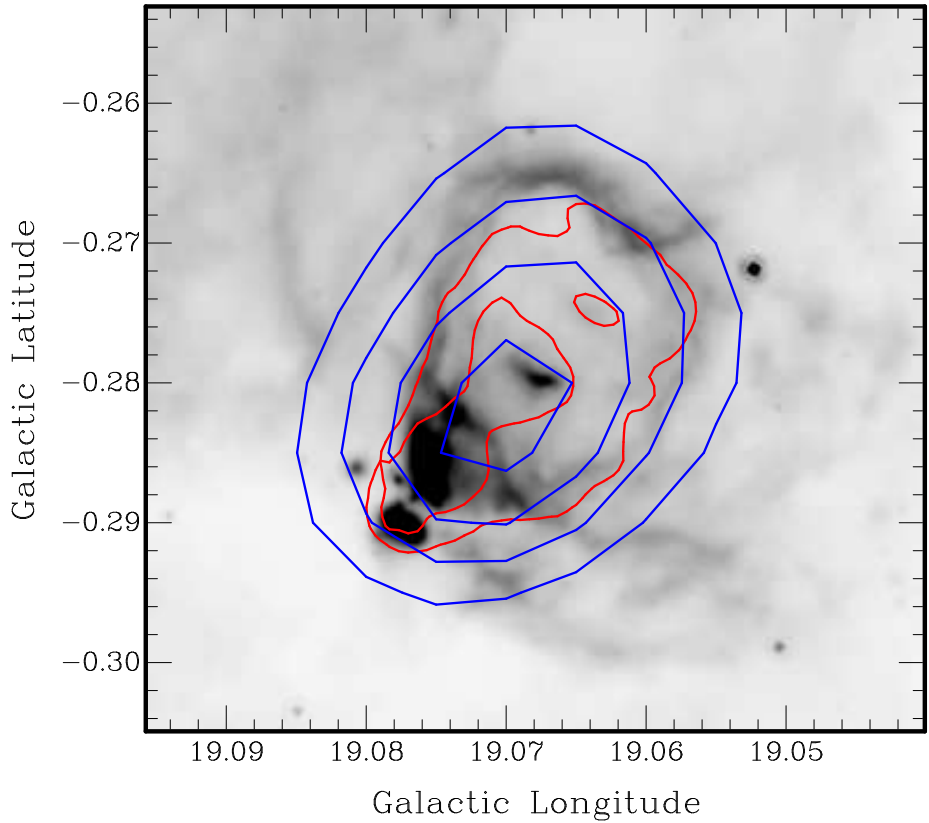}
\caption{The G19.07-0.28 \ion{H}{2} region. The bubble structure of
  this \ion{H}{2} region is clearly shown in the GLIMPSE 8.0 $\mu$m
  image. The red (or lighter) contour (at 950 MJy sr$^{-1}$) traces 24
  $\mu$m emission from hot dust filling the interior of the
  bubble. The 24~$\mu$m emission is saturated in the MIPSGAL image
  within the innermost contour shown. Radio continuum emission at 1420
  MHz from the VGPS (blue/darker contours; 100 to 175 K at 25 K
  intervals) also peaks within the infrared bubble. \label{fig:bubble}}
\end{figure}

The higher resolution GLIMPSE 8.0~$\mu$m image of the region is shown
in Figure~\ref{fig:bubble}. At this wavelength the source is clearly a
bubble structure with a radius of $\sim 0\fdg 01$
($\sim0.8$~pc). Ionized gas, traced by the 1420 MHz radio emission,
and warm dust, traced by 24~$\mu$m emission, fill the central
region. The compact bubble morphology of the region and the agreement
in the stellar spectral type determined by the radio and infrared flux
support the idea that the region is ionization bounded ($f\sim1$).

\subsection{G18.88-0.49}

The radio source G18.88-0.49 is the second brightest of the compact radio
sources found on the periphery of W~39 with $F_\nu=1.53\pm0.02$~Jy.
From this we find $\log\left(fQ\right)=48.45$ equivalent to a single O8~V
star. The ionized gas mass is $M\sim 44$~M$_\sun$ and the average
electron density is $n_e\sim160$~cm$^{-3}$.

\begin{figure}
\plotone{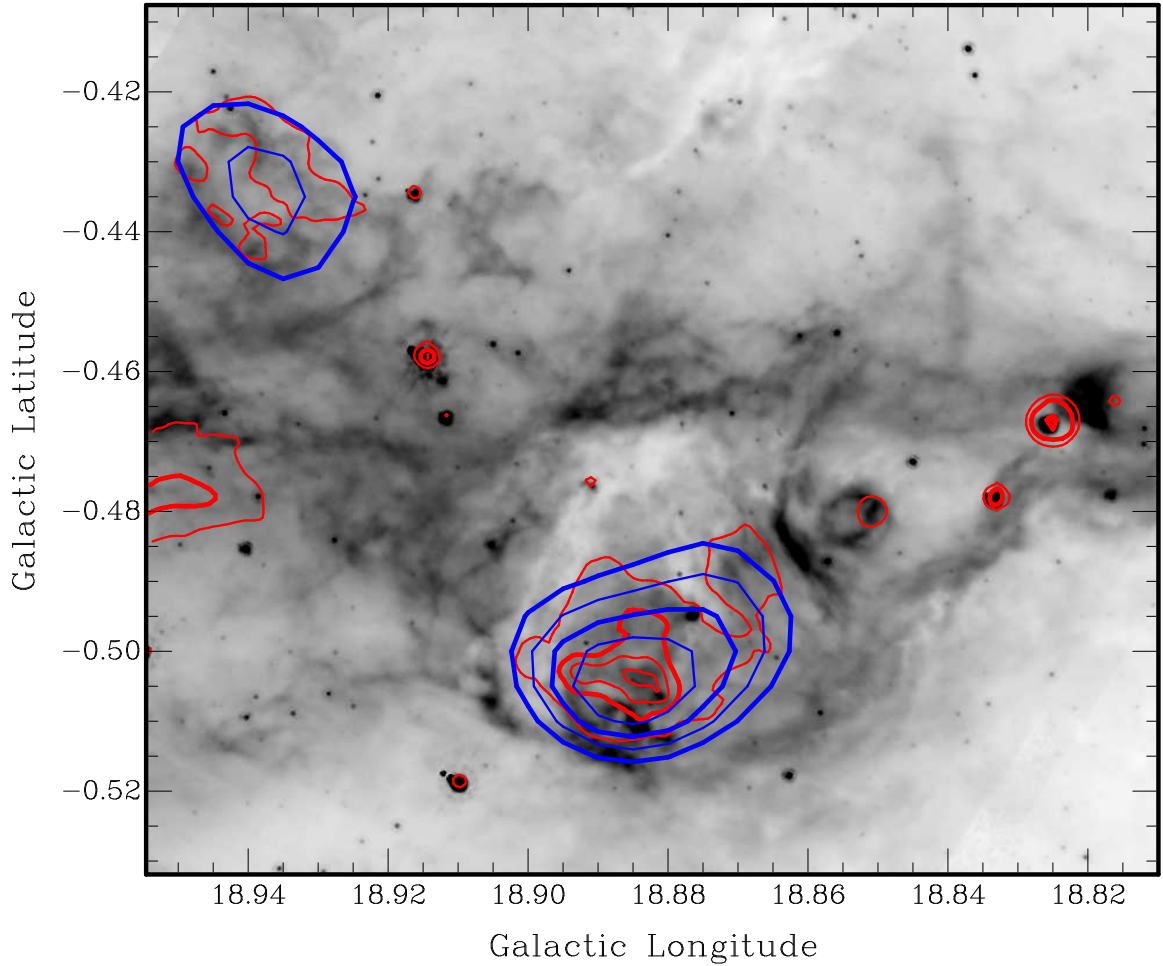}
\caption{The \ion{H}{2} regions G18.88-0.49 and G18.94-0.43. This GLIMPSE
  8.0~$\mu$m image shows two of the compact \ion{H}{2} regions
  surrounding W~39. Red (or lighter) contours (at 550, 1000, and 1550 MJy
  sr$^{-1}$) trace MIPSGAL 24 $\mu$m emission, and blue (or darker)
  contours trace 1420 MHz radio continuum emission (80
  to 110 at 10 K intervals). Note that the MIPSGAL image is saturated
  at the center of the G18.88-0.49 \ion{H}{2} region.  \label{fig:west}}
\end{figure}

Unlike G19.07-0.28 there is no IRAS point source associated with this
region. The higher resolution GLIMPSE 8.0~$\mu$m image of the region
is shown in Figure~\ref{fig:west}. Although the morphology is more
complex than G19.07-0.28, the presence of bright-rimmed structures at
the edge of the radio continuum emission along with the coincident
24~$\mu$m emission suggest that we are also looking at a dusty,
ionization-bounded, bubble structure similar in size to G19.07-0.28.

\subsection{G18.94-0.43}

The remaining radio source on the periphery of W~39 is G18.94-0.43. We
measured $F_\nu=0.8\pm0.09$~Jy, which corresponds to
$\log\left(fQ\right)=48.16$, equivalent to a single O9~V star. The
ionized gas mass is $M\sim 32$~M$_\sun$, and the average electron
density is $n_e\sim110$~cm$^{-3}$.

As with G18.88-0.49 there is no IRAS point source associated with the
region. The high resolution image of the region shown in
Figure~\ref{fig:west} reveals a complex morphology that is difficult
to interpret in detail. As we would expect for an \ion{H}{2} region
there appears to be bright-rimmed structures along the low-latitude portion
of the region and a minimum in the 8~$\mu$m emission coincident with
the radio continuum maximum.

\subsection{Timescales \& Triggered Star Formation} 

The radio continuum view of the W~39 region is of an evolved
massive HII region (powered by multiple mid- to early-type O stars)
with multiple areas of less intense massive star formation (i.e.,
the three \ion{H}{2} regions powered by single late O stars) occurring along the periphery of the
region. One possibility is that these secondary regions of star
formation have been triggered by the expansion of the W~39 \ion{H}{2}
region. While proving this idea is difficult we can at least
investigate the relevant timescales for the most clearly defined
secondary region, G19.07-0.28, to see if this scenario is at all
plausible.

We can obtain a rough estimate of the age of G19.07-0.28 using the
classic description of the evolution of an \ion{H}{2} region
\citep{spi78, ost89}. The O star first creates a ionized gas region about
the size of the Str\"{o}mgren sphere on an essentially instantaneous
timescale of $\sim 10^5/n_e{\rm (cm^{-3})}$~yr. The overpressurized
\ion{H}{2} region then expands from its initial radius, $r_s$ to the
observed radius $R$ at time $t$ as,
\begin{equation} \label{eqn:h2evo}
\frac{R}{r_s} = \left(1+\frac{7 C_{II} t}{4r_s}\right)^{4/7} ,
\end{equation}
where $C_{II}$ is the isothermal sound speed in the ionized gas ($\sim
10$~km~s$^{-1}$).

To estimate the initial density of the cloud in which the \ion{H}{2}
region formed we examined the nearby 1.1 mm source from the BGPS,
G019.077-00.287.  The peak brightness of this source is 2.3 Jy
beam$^{-1}$ corresponding to a column density of $N_{H_2} =
7.9\times10^{22}$~cm$^{-2}$ using equation A.27 in \citet{kau08}.
The deconvolved radius of the source is $93\farcs82$ ($\sim
2$~pc). Adopting a path length of 4~pc we obtain a density
$n\sim10^{3.8}$~cm$^{-3}$.  In such an environment an O8~V star would
rapidly create an \ion{H}{2} region of size $r_s \sim 0.2$~pc. Using
this value and the observed size of the region $\sim 0.8$~pc in
equation~\ref{eqn:h2evo} we find $t\sim10^5$ years. 

While this is clearly a rough estimate it is important to note
that this timescale is approximately one order of magnitude smaller
than the timescales appropriate for the evolution of the much larger
W~39 region. As pointed out by \citet{chu75}, an approximate lower
limit to the age of an evolved \ion{H}{2} region can be obtained by
the sound crossing time, $\sim 1.5\times10^6$ years in the case of W~39. The lack of
non-thermal radio emission from SNRs gives a consistent
upper limit to the age of $2-4\times10^6$ years based on the main
sequence lifetime of early O stars \citep{sch97}. We conclude that the
probable timescales associated with the expansion of W~39 ($\sim 10^6$
years) and the likely ages of the secondary regions of massive star formation
($\sim 10^5$ years)  are consistent with a scenario of triggered star
formation.

\section{Star Formation Traced by MIPSGAL \& GLIMPSE}\label{sec:yso}

\subsection{GLIMPSE Detected YSO Candidates}

\begin{figure}
\plotone{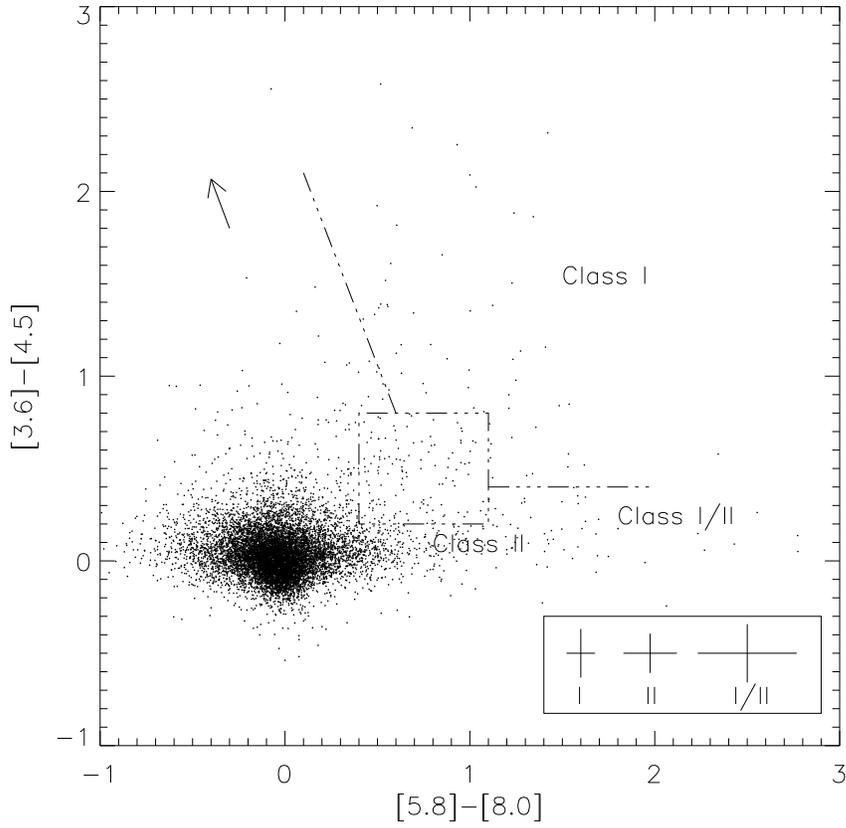}
\caption{GLIMPSE Color-Color Diagram. A total of 10657
sources with valid magnitudes in all four IRAC bands and located in
the W~39 field (see text) are plotted. The YSO identification criteria
from \citet{all04} are shown as dash-dot lines. For clarity only the
median color errors are shown for each of the YSO classes. A reddening
vector for a YSO with $A_V=20$ is shown from \citet{meg04}. \label{fig:ccd}}
\end{figure}

In order to determine if the only recent, and thus likely triggered,
star formation activity surrounding W~39 is restricted to the three
compact \ion{H}{2} regions discussed previously, we used the
\emph{Spitzer} GLIMPSE catalog to identify YSO candidates (YSOs
hereafter) in the W~39 field  within $18\fdg5 < l < 19\fdg3$, $-0\fdg55
< b < -0\fdg1$. The high latitude cutoff was chosen to avoid
including the prominent infrared dark clouds (IRDCs) found at
$b>-0\fdg1$ which are not associated with W~39. A total of 10657
sources with valid magnitudes in all four IRAC bands were
identified. Figure~\ref{fig:ccd} is a color-color (CC) diagram showing
all of these sources along with the YSO classification criteria from
\citet{all04}. We identify 57 Class I YSOs, 129 Class II YSOs and 57
Class I/II YSOs ([3.6]-[4.5] $<0.4$ and [5.8]-[8.0]$>1.1$). Class I
YSOs are objects with a substantial circumstellar envelope, Class II
YSOs have most of the circumstellar material residing in a disk, and
Class I/II objects are most likely Class II or stellar objects with
poor [8.0] photometry causing them to scatter into this region of the
CC diagram. In the canonical picture, a Class I YSO will evolve to a
Class II YSO with each stage of evolution lasting for $\tau_I \sim
10^5$ years and $\tau_{II} \sim 10^6$ years respectively
\citep{war02}. 

\begin{figure}
\plotone{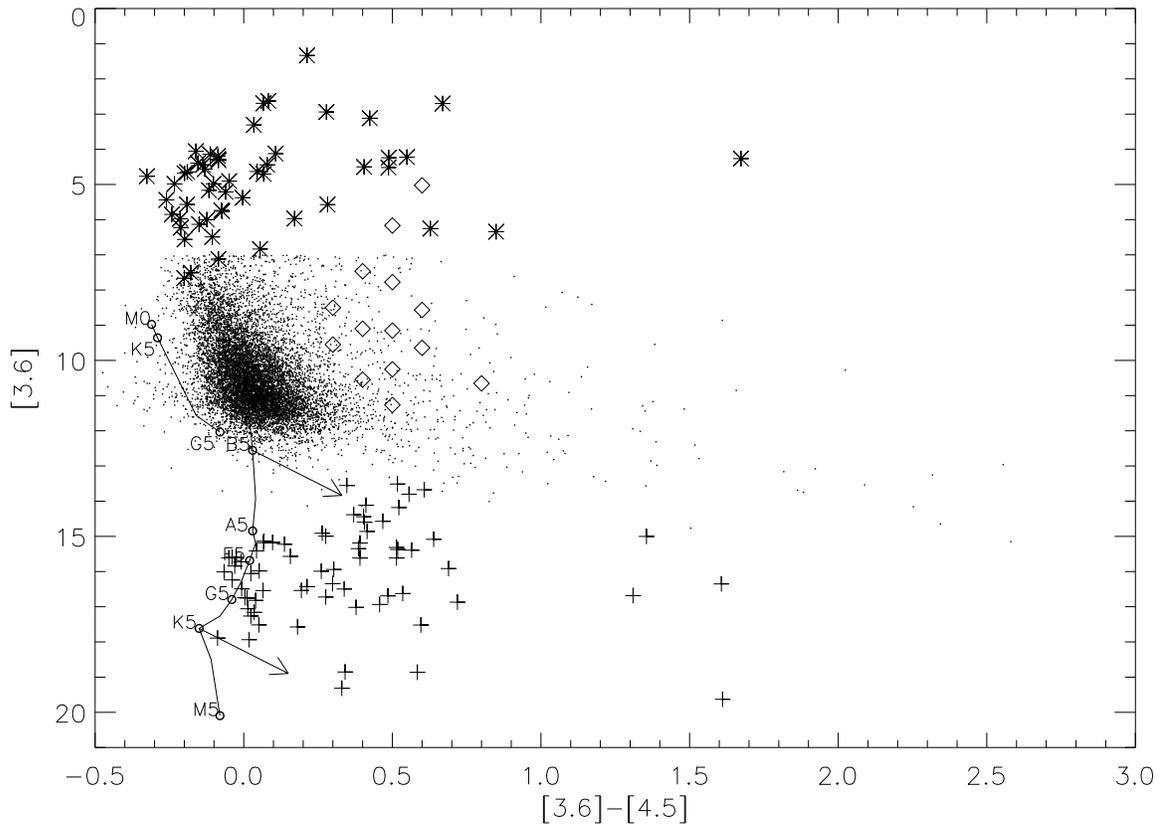}
\caption{GLIMPSE Color-Magnitude Diagram.  The sample from
  Figure~\ref{fig:ccd} is plotted along with a main sequence and giant
  branch shifted to the distance of W~39.  Note the bright
  limit at $\sim 7$ due to saturation and a faint, color-dependent,
  sensitivity cutoff at [3.6] $\sim 13-15$. Also shown are
  representative T Tauri stars (crosses), Herbig AeBe stars (diamonds)
  and AGB stars (asterisks) also plotted at a distance of 4.5 kpc (see text
  for details). The CMD shows that, at the distance of W~39, our
  GLIMPSE sample will be able to detect  intermediate-mass YSOs, and
  that the primary source of confusion for red objects will be
  background AGB stars. \label{fig:cmd}}
\end{figure}

In Figure~\ref{fig:cmd} the sample is plotted in a  [3.6]-[4.5] vs. [3.6] color-magnitude
(CM) diagram. We see there is a bright cutoff
at [3.6] $\sim 7$ due to saturation and a faint, color-dependent,
sensitivity cutoff at [3.6] $\sim 13-15$. Included in the CM diagram
are two lines showing the position of main-sequence stars (B0~V to
M5~V) and giant stars (G5~III to M0~III) at a distance of 4.5 kpc. The
crosses are T-Tauri stars from \citet{har05} shifted from the distance
of Taurus-Auriga to W~39. The diamonds are intermediate-mass Herbig AeBe
stars from \citet{the94} and \citet{fin84} again shifted to the
distance of W~39. For the Herbig AeBe sample we have used L and M band
magnitudes as rough proxies for [3.6] and [4.5]. Finally, the asterisks
represent AGB stars from the \emph{Spitzer}/IRAC study of Galactic AGB
stars by Reiter, Marengo, \& Fazio (in preparation). The CM
diagram illustrates that our sample is primarily sensitive to
intermediate-mass YSOs at a distance of 4.5 kpc. The CM diagram also
provides some insights into likely contaminants in our sample. We see
that the majority of background main-sequence stars are too faint to
be detected while foreground main-sequence stars with moderate amounts
of reddening will be a source of contamination for the least-red Class
II YSOs. The major contaminant, for both Class I and Class II YSOs, are
reddened background AGB stars. Unfortunately, since we are looking
toward the inner Galaxy, background stellar distances approaching 20 to
30 kpc are possible meaning even the faintest, reddest, sources in our
sample could be matched by a distant, heavily reddened AGB.

\begin{figure}
\epsscale{0.75}
\plotone{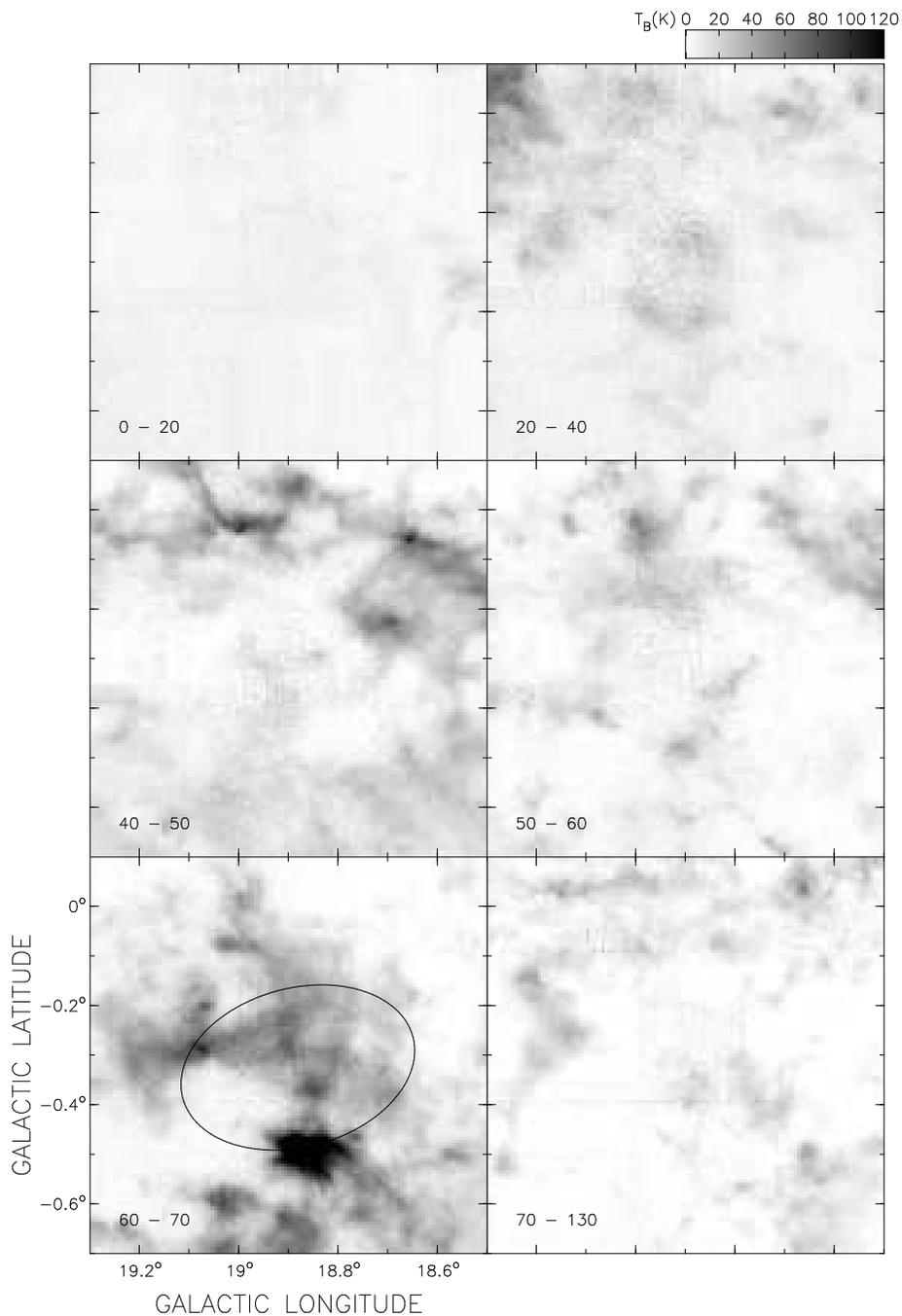}
\caption{Integrated $^{13}$CO GRS images between $0 \leq V_\mathrm{LSR} \leq
+130$~km~s$^{-1}$. The velocity increments are shown in the lower-left
corner of each panel, and the greyscale is the same in all panels. Note
the similarity of the emission seen at high latitude in the $40-50$ km~s$^{-1}$
panel and the IRDCs seen in Figure~\ref{fig:ir8_24}, and the concentration of
material associated with the two main regions of star formation
around W~39 shown in the $60-70$ km~s$^{-1}$ panel. The ellipse
is the same as that shown in Figure~\ref{fig:w39cont}. \label{fig:co}}
\epsscale{1.0}
\end{figure}

\begin{figure}
\plotone{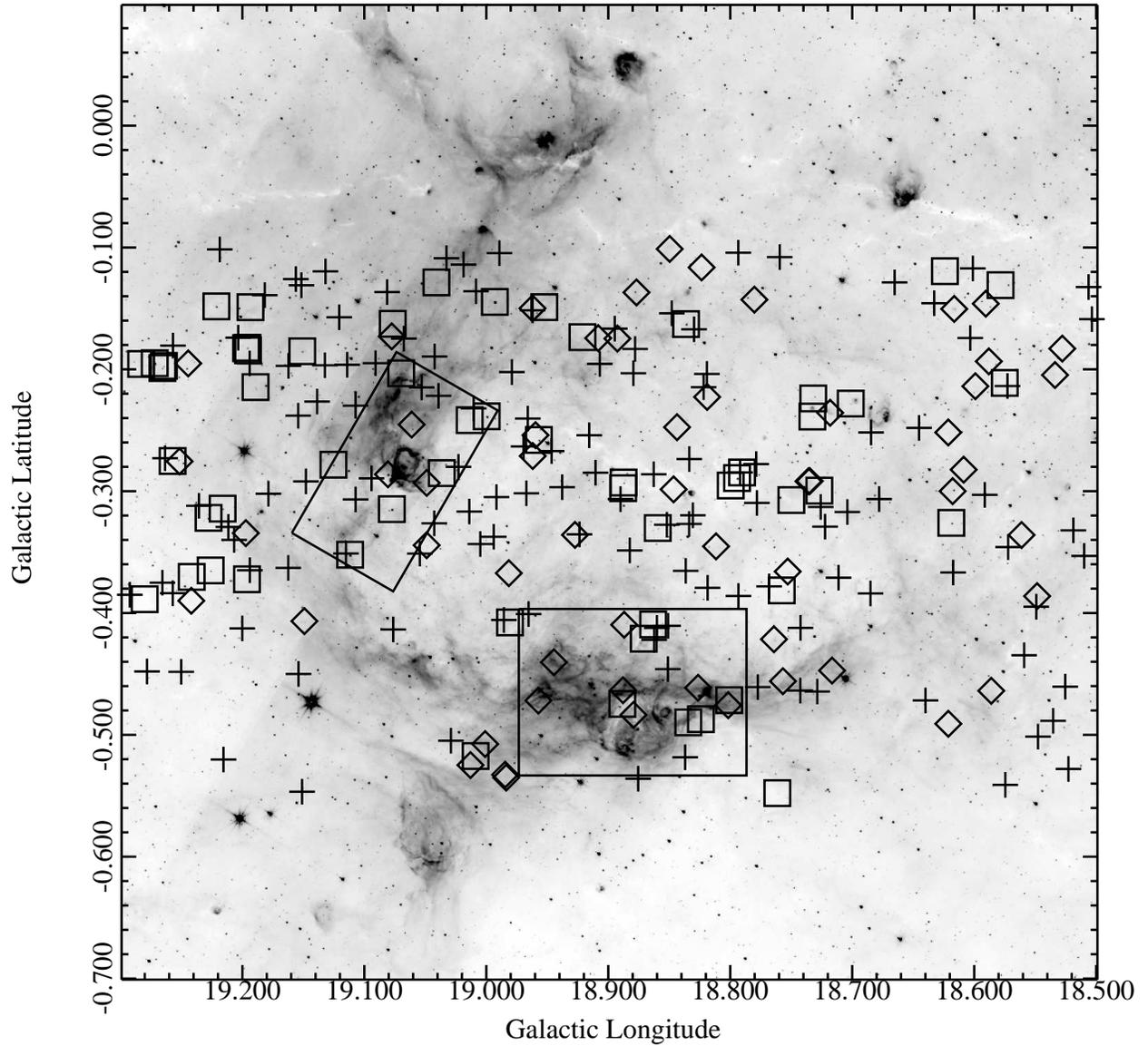}
\caption{W~39 YSO candidates. The location of the various YSO
  candidates identified using the GLIMPSE CC diagram
  (Figure~\ref{fig:ccd}) are shown: Class I = square, Class II = +,
  Class I/II = diamond. The rectangles indicate regions around
  G18.88-0.49 and G19.07-0.28 where the YSO population was
  investigated in more detail. \label{fig:pos}}
\end{figure}

In Figure~\ref{fig:pos} we show the spatial distribution of the
various YSO classes. It is clear that star formation activity
is not uniformly distributed along the periphery of the W~39 bubble
with the most striking feature being the lack of YSOs at the
high and low longitude ends of the bubble around $b=-0\fdg15$ and
$b=-0\fdg45$ respectively. This is consistent with the lack of
molecular gas seen at these locations in the $^{13}$CO images shown in
Figure~\ref{fig:co}. There is also a swath of YSOs located within the
infrared bubble that roughly follows the $^{13}$CO distribution seen in
the lower left panel of Figure~\ref{fig:co}. Almost all of the $\sim
10$ Class I sources found within the bubble are associated with an
IRDC that runs roughly from $l=19\fdg0$ to $l=18\fdg8$ at
$b\sim-0\fdg3$. If associated with W~39, these sources could represent
recent star formation occurring in the near or far wall of the bubble,
but it is also possible that the IRDC is not directly related to
W~39. We tried to measure the $V_\mathrm{LSR}$ of the IRDC, but were
unable to clearly identify any $^{13}$CO features that were
unambiguously associated with the IRDC. Higher resolution radio/mm
line observations of this IRDC in tracers of cold, dense gas would
help to clarify its relation to W~39. 

\begin{figure}
\plotone{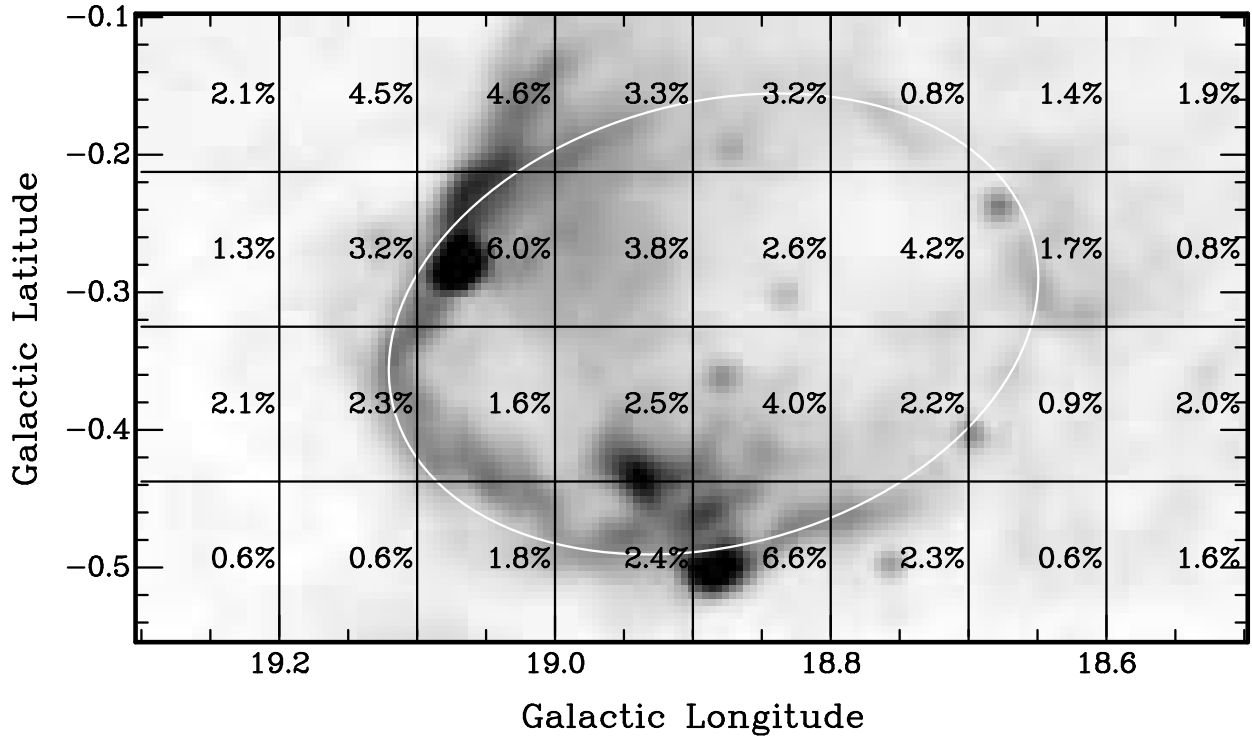}
\caption{YSO Candidate Frequency. The percentage of GLIMPSE sources in
  each $\sim 0\fdg1 \times 0\fdg1$ area classified as YSO candidates is
  shown. For reference the greyscale shows VGPS 1420 MHz emission and the
  white ellipse indicates the extent of the IR bubble. Note that
  the regions around G18.88-0.49 and G19.07-0.28 have the highest YSO
  percentage in the field and stand well above the median background
  value of 1.8\%, which was defined using the eight high and low longitude
  areas. \label{fig:red}}
\end{figure}

Qualitatively it appears from Figure~\ref{fig:pos} that most of the
star-formation activity located on the periphery of W~39 is
concentrated around the compact \ion{H}{2} regions. In order examine
this quantitatively we divided our search area into 32 roughly $0\fdg1
\times 0\fdg1$ areas and calculated the percentage of sources in each
area that have a YSO classification (see Figure~\ref{fig:red}). The
``background'' YSO percentage, using the median of the eight high and
low longitude columns, is 1.8\% with an interquartile range of
0.85\%. The YSOs within the bubble show up as a slight excess over the
background ($\sim 4$\%) but the main outliers are the areas around
G18.88-0.49 and G19.07-0.28 with YSO percentages of 6.6\% and 6.0\%
respectively.

\subsection{MIPSGAL Detected YSOs around G18.88-0.49 and G19.07-0.28}

We investigated the YSO population around G18.88-0.49 (19 YSOs; G18 hereafter) and
G19.07-0.28 (17 YSOs; G19 hereafter) in more detail (the areas are
shown in Figure~\ref{fig:pos}). For each YSO, JHK
photometry was obtained either from 2MASS or UKIDSS, and 24 $\mu$m
photometry was obtained from MIPSGAL images if possible. UKIDSS
photometry was used if an unambiguous match to the IRAC source could
be made using a $0\farcs6$ search radius, otherwise the associated
2MASS photometry given in the GLIMPSE catalog was used. Aperture
photometry on the MIPSGAL images was obtained for 15 of the YSOs using
a 6 arcsec radius aperture, a user-defined aperture correction of
1.67, and a zero-magnitude flux density of 7.17 Jy \citep{eng07}. We
have adopted a conservative uncertainty of 10\% for the [24]
photometry due to image background variations and uncertainty in the
aperture correction estimate. The YSO sample photometry is summarized
in Table~\ref{tbl:ysophot}.

\begin{figure}
\plotone{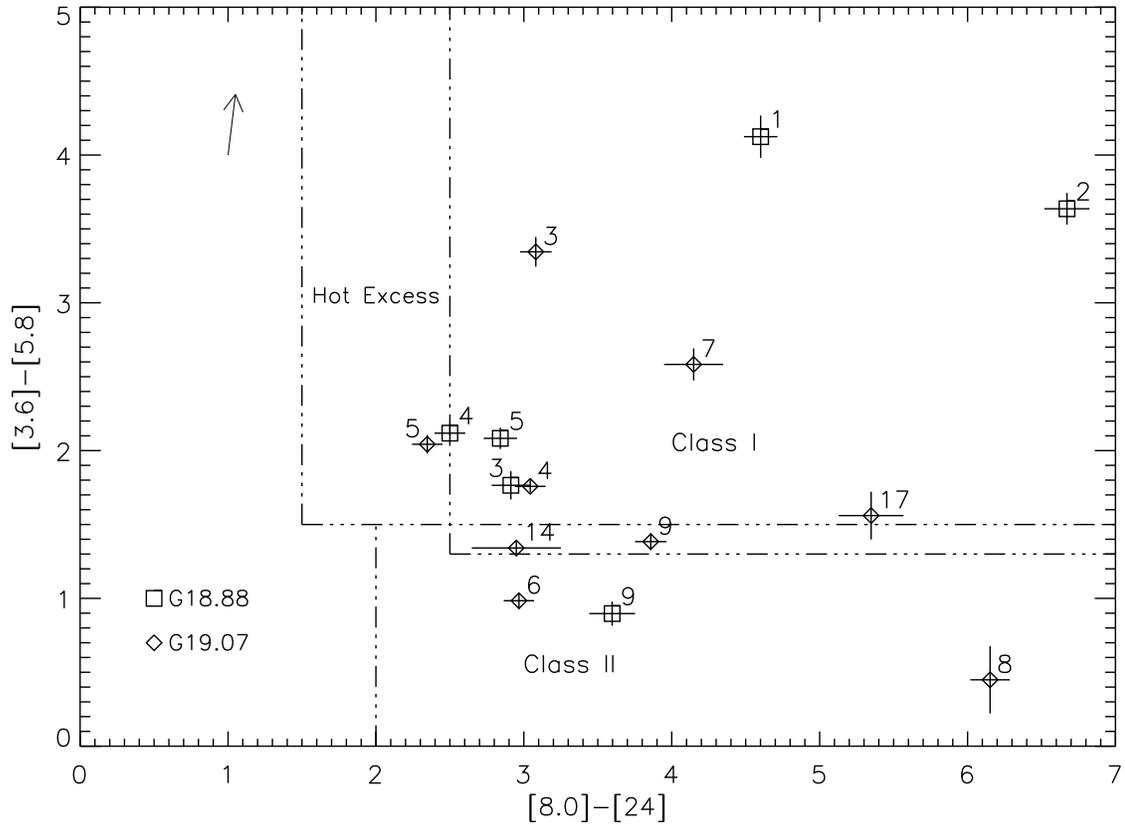}
\caption{GLIMPSE and MIPSGAL Color-Color Diagram. YSOs in the G18.88-0.49 and
  G19.07-0.28 regions (the rectangles in Figure~\ref{fig:pos}) with 24 $\mu$m
  photometry are plotted along with the YSO classification criteria of
  \citet{rho06}. A reddening vector for $A_V=20$ from \citet{rho06} is
  shown. \label{fig:ccd13_4}}
\end{figure}

When available the 24 $\mu$m photometry provides a very clear
discriminant between Class I and Class II YSOs. At or longward of 24
$\mu$m a Class II YSO's spectral energy distribution (SED) is
typically flat or falling while a Class I YSO's SED will typically be
rising. In Figure~\ref{fig:ccd13_4} we plot the sample of YSOs with
24~$\mu$m photometry on a [3.6]-[5.8] vs. [8.0]-[24] CC diagram along
with the YSO classification criteria from \citet{rho06}. As can be seen from
comparing columns two and three of Table~\ref{tbl:ysoclass} the
IRAC-only and IRAC \& MIPS YSO classifications are typically in
agreement. Two of the YSOs are classified in the \citet{rho06} scheme
as possible ``hot excess'' sources, which are either very heavily
embedded Class II YSOs or Class I YSOs undergoing very rapid accretion.

To confirm that our YSOs are most likely intermediate mass objects
(i.e., $\sim 2-10$~M$_\sun$) we obtained YSO model fits
using the SED fitting program of \citet{rob07} for those sources with
24 $\mu$m photometry. Following \citet{ale12} we use apertures of
2\arcsec\ for UKIDSS, 3\arcsec\ for 2MASS and GLIMPSE, and 7\arcsec\ for
MIPSGAL and set a minimum flux uncertainty of 10\% to prevent any one
data point from dominating the fit. We allowed the distance to vary
between 4.3 and 4.7 kpc and $A_V$ to vary between 5 and 50
magnitudes. In Table~\ref{tbl:modfit} we show the details
of the best-fitting YSO model as well as the YSO stage, where Stage I
has $\dot{M_e}/M_* > 10^{-6}$ yr$^{-1}$, Stage II has $\dot{M_e}/M_* <
10^{-6}$ yr$^{-1}$ and $M_d/M_* > 10^{-6}$ and Stage III has
$\dot{M_e}/M_* < 10^{-6}$ yr$^{-1}$ and $M_d/M_* < 10^{-6}$
($\dot{M_e}$ is the envelope accretion rate in $M_\odot$ yr$^{-1}$,
$M_*$ and $M_d$ are the stellar and disk masses respectively in
$M_\odot$), and the stages typically correspond one-to-one with the
observationally defined Class I, II, and III. 

The SED fitter typically returned multiple models that were equally
good fits statistically. As outlined in \citet{ale12} there are
techniques that can be used to obtain a weighted average model for the
YSO based on this redundancy. In this paper, since we are using the
SED fitter primarily to compare with CC diagram classifications and as
a rough check on the YSO mass, we simply report the details of the
single best-fitting model. We find that our best-fit models are all in
the intermediate-mass range $\sim 2-10$~M$_\sun$. 

A comparison between the SED fitting results and the results from the
\citet{rho06} CC diagram is interesting. Sources G18-1, G18-2, G19-3,
and G19-7 all lie well within the Class I region of the CC diagram and
are also identified as Stage I objects by the SED fitter. All but one
of the YSOs with [3.6]-[5.8] between 1.5 and 2.5, including the two
``hot excess'' sources, are classified as reddened Stage II objects by
the SED fitter. The exception is the reddest source, G19-17 which is
classified as Stage I. For the sources with [3.6]-[5.8]$<1$ agreement
between the CC diagram and the SED fitter is mixed; the SED fitter
classifies G19-6 as Stage II, but classifies G18-9 as Stage III and
the very red G19-8 YSO as Stage I.

\section{Conclusions}\label{sec:conc}

This study represents the first study of W~39 at high resolution at
infrared and radio wavelengths. The main conclusions of our study are:

1) W~39 is a giant \ion{H}{2} region powered by numerous OB
stars with a likely total ionizing luminosity $\log(Q) > 50$, roughly
equivalent to a cluster of over 30 O8 V stars. Future work will focus
on identifying the OB stars powering W~39, using techniques described in
\citet{ale12}, as the first step toward obtaining a non-kinematic
distance estimate to the region.

2) Three compact \ion{H}{2} regions are located on the periphery of
W~39. These regions are each likely powered by a single late O-type star (O7
- O9 V). Two of the regions have a distinct dust-filled bubble
morphology seen in infrared 8.0 and 24 $\mu$m images.

3) Simple kinematic models suggest that the Myr timescales associated
with the expansion of W~39 are about an order of magnitude larger than
the evolutionary timescales associated with the smaller \ion{H}{2} regions. This is
consistent with the scenario where the formation of the smaller
\ion{H}{2} regions is triggered by the expansion of W~39.

4) Recent star formation surrounding W~39 is not restricted just to
the massive stars powering the the small \ion{H}{2} regions. We used
infrared data to identify a number of young/heavily-embedded
intermediate-mass YSOs not directly associated with the \ion{H}{2}
regions.

5) Star formation activity is not distributed uniformly around the
periphery of W~39, rather it is concentrated in the two regions
surrounding the small \ion{H}{2} regions and potentially in an IRDC
that may be associated with the front or rear wall of a bubble
structure. Future studies of the W~39 region will explore the stellar
content of the compact \ion{H}{2} regions in more detail, and will
examine the potential relationship of the IRDC with W~39. 

\acknowledgments
This work is partially supported through NASA grant
ADAP-NNX10AD55G. The authors would like to thank Massimo Marengo for
sharing the {\it Spitzer} AGB data and Chip Kobulnicky for his helpful
comments and suggestions regarding YSO identification.

{\it Facilities:} \facility{Spitzer}, \facility{2MASS},
\facility{UKIRT}, \facility{CSO}, \facility{VLA}, \facility{FCRAO}



\clearpage 

\clearpage 

\clearpage 

\clearpage

\clearpage

\clearpage

\clearpage

\clearpage

\clearpage

\clearpage

\clearpage


\clearpage

\begin{deluxetable}{llcccccccccc}
\tabletypesize{\scriptsize}
\rotate
\tablecaption{Photometry for YSO Candidates in the G18.88 and G19.07 Regions \label{tbl:ysophot}}
\tablewidth{0pt}
\tablehead{
\colhead{ID} & \colhead{SSTGLMC} & \colhead{J} & \colhead{H} & \colhead{K$_s$} & \colhead{[3.6]} & \colhead{[4.5]} & \colhead{[5.8]} & \colhead{[8.0]} & \colhead{[24]}
}
\startdata
G18-1  & G018.8008-00.4714 & 18.621(0.073) & 16.716(0.028) & 15.152(0.018) & 13.256(0.134) & 10.939(0.044) &  9.132(0.050) &  7.711(0.053) & 3.11(0.1) \\
G18-2  & G018.8885-00.4746 & 17.940(0.04)  & 16.828(0.035) & 14.651(0.013) & 12.963(0.083) & 10.408(0.199) &  9.327(0.068) &  9.402(0.115) & 2.73(0.1) \\
G18-3  & G018.8709-00.4210 & 17.628(0.03)  & 15.308(0.009) & 13.312(0.004) & 10.994(0.071) & 10.034(0.068) &  9.229(0.064) &  8.482(0.082) & 5.57(0.1) \\
G18-4  & G018.8240-00.4871 & \nodata       & \nodata       & 14.615(0.011) & 11.293(0.049) &  9.951(0.054) &  9.175(0.047) &  8.480(0.028) & 5.98(0.1) \\  
G18-5  & G018.8632-00.4093 & \nodata       & \nodata       & 13.420(0.004) & 10.911(0.058) &  9.817(0.054) &  8.828(0.043) &  7.951(0.052) & 5.11(0.1) \\ 
G18-6  & G018.8344-00.4897 & \nodata       & \nodata       & \nodata       & 13.080(0.146) & 11.157(0.090) &  9.890(0.101) &  9.392(0.179) & \nodata   \\
G18-7  & G018.9653-00.4008 & \nodata       & 15.743(0.013) & 13.357(0.004) & 12.016(0.187) & 11.761(0.094) & 11.244(0.120) & 10.646(0.149) & \nodata   \\
G18-8  & G018.8509-00.4104 & 14.680(0.041) & 13.260(0.035) & 12.293(0.036) & 10.645(0.048) & 10.104(0.061) &  9.594(0.055) &  8.950(0.085) & \nodata   \\
G18-9  & G018.8623-00.4220 & 18.260(0.06)  & 14.051(0.079) & 11.900(0.034) & 10.366(0.053) & 10.160(0.044) &  9.469(0.062) &  8.657(0.118) & 5.06(0.1) \\
G18-10 & G018.8601-00.4122 & \nodata       & \nodata       & \nodata       & 11.964(0.066) & 11.417(0.094) & 10.627(0.089) &  9.892(0.137) & \nodata   \\
G18-11 & G018.8369-00.5183 & \nodata       & 16.546(0.024) & 13.760(0.056) & 11.882(0.081) & 11.451(0.070) & 10.882(0.072) & 10.440(0.154) & \nodata   \\
G18-12 & G018.8510-00.4460 & 17.477(0.026) & 13.779(0.002) & 11.874(0.001) & 10.557(0.036) & 10.268(0.065) &  9.792(0.070) &  9.337(0.165) & \nodata   \\
G18-13 & G018.8793-00.4835 & \nodata       & 17.544(0.067) & 15.021(0.018) & 13.201(0.121) & 13.101(0.154) & 10.617(0.226) &  9.372(0.305) & \nodata   \\
G18-14 & G018.8879-00.4633 & 14.138(0.038) & 12.501(0.028) & 11.897(0.025) & 11.399(0.083) & 11.408(0.054) & 11.528(0.339) &  9.581(0.274) & \nodata   \\
G18-15 & G018.9445-00.4404 & 16.538(0.012) & 14.426(0.004) & 13.203(0.004) & 12.248(0.131) & 12.187(0.124) & 11.402(0.237) & 10.203(0.199) & \nodata   \\
G18-16 & G018.8869-00.4096 & 18.539(0.07)  & 15.156(0.008) & 13.589(0.005) & 12.613(0.072) & 12.436(0.099) & 12.080(0.197) & 10.820(0.159) & \nodata   \\
G18-17 & G018.8263-00.4613 & \nodata       & 17.650(0.07)  & 14.761(0.01)  & 12.656(0.117) & 12.484(0.169) & 10.700(0.139) &  9.203(0.223) & \nodata   \\
G18-18 & G018.8016-00.4756 & 13.295(0.033) & 12.535(0.033) & 12.282(0.039) & 12.113(0.079) & 12.062(0.065) & 11.586(0.334) &  8.812(0.250) & \nodata   \\
G18-19 & G018.9573-00.4718 & \nodata       & 12.775(0.046) & 11.934(0.040) & 11.152(0.135) & 11.092(0.123) & 10.313(0.164) &  8.804(0.350) & \nodata   \\ 
G19-1  & G019.0697-00.2032 & \nodata       & \nodata       & 14.885(0.015) & 12.787(0.135) & 12.278(0.155) & 10.854(0.300) &  9.324(0.141) & \nodata   \\
G19-2  & G019.0365-00.2851 & \nodata       & \nodata       & 15.916(0.042) & 12.937(0.118) & 12.195(0.111) & 11.342(0.126) & 10.236(0.102) & \nodata   \\
G19-3  & G019.0136-00.2416 & \nodata       & \nodata       & \nodata       & 13.536(0.088) & 11.447(0.084) & 10.191(0.048) &  9.191(0.038) & 6.11(0.1) \\
G19-4  & G019.1116-00.3521 & 16.085(0.008) & 12.961(0.001) & 10.335(0.001) &  7.614(0.030) &  6.704(0.051) &  5.856(0.029) &  5.093(0.028) & 2.05(0.1) \\
G19-5  & G019.0774-00.3147 & \nodata       & 15.291(0.009) & 11.974(0.001) &  8.202(0.052) &  7.078(0.059) &  6.159(0.035) &  5.557(0.025) & 3.21(0.1) \\
G19-6  & G019.1253-00.2782 & 13.528(0.001) & 12.207(0.001) & 10.870(0.001) &  9.380(0.039) &  8.908(0.047) &  8.396(0.038) &  7.057(0.025) & 4.09(0.1) \\
G19-7  & G018.9996-00.2385 & \nodata       & 13.627(0.036) & 11.308(0.046) &  8.863(0.090) &  7.253(0.102) &  6.280(0.061) &  5.708(0.173) & 1.56(0.1) \\
G19-8  & G019.0391-00.2217 & 15.994(0.007) & 13.901(0.003) & 12.665(0.002) & 11.909(0.061) & 11.638(0.080) & 11.460(0.219) & 10.402(0.088) & 4.25(0.1) \\
G19-9  & G019.0530-00.2148 & 15.734(0.006) & 13.040(0.001) & 10.950(0.001) &  9.115(0.039) &  8.367(0.095) &  7.731(0.043) &  6.798(0.036) & 2.94(0.1) \\
G19-10 & G019.0103-00.2320 & 17.523(0.027) & 14.303(0.004) & 12.552(0.002) & 11.372(0.056) & 11.070(0.073) & 10.586(0.114) & 10.047(0.150) & \nodata   \\
G19-11 & G019.1071-00.3067 & \nodata       & 14.602(0.005) & 12.529(0.002) & 11.250(0.087) & 10.965(0.098) & 10.803(0.094) & 10.327(0.075) & \nodata   \\
G19-12 & G019.0941-00.2894 & 16.036(0.008) & 13.720(0.002) & 12.085(0.002) & 10.382(0.034) &  9.651(0.045) &  9.012(0.045) &  8.423(0.076) & \nodata   \\
G19-13 & G019.0228-00.2800 & \nodata       & 14.449(0.004) & 12.779(0.003) & 11.545(0.064) & 11.302(0.090) & 11.188(0.129) & 10.597(0.249) & \nodata   \\
G19-14 & G019.1150-00.3511 & 14.682(0.003) & 11.561(0.001) &  9.568(0.001) &  7.371(0.044) &  6.846(0.041) &  6.030(0.031) &  5.449(0.026) & 2.5(0.3)  \\
G19-15 & G019.0488-00.2932 & 13.602(0.001) & 12.428(0.001) & 11.863(0.001) & 11.381(0.068) & 11.312(0.089) & 10.571(0.125) &  9.346(0.140) & \nodata   \\
G19-16 & G019.0806-00.2861 & 17.782(0.04)  & 15.623(0.01)  & 14.229(0.009) & 11.043(0.121) & 11.270(0.184) &  8.043(0.113) &  6.651(0.286) & \nodata   \\
G19-17 & G019.0612-00.2453 & 15.077(0.004) & 13.922(0.003) & 13.219(0.004) & 12.274(0.083) & 11.964(0.123) & 10.714(0.139) &  9.128(0.195) & 3.78(0.1) \\
\enddata
\tablecomments{JHK photometry is from 2MASS for Sources G18-8, G18-9 (H,K), G18-11(K), G18-14, G18-18, G18-19, G19-7, and from UKIDSS-GPS for others.}
\end{deluxetable}

\clearpage 

\begin{deluxetable}{lccc}
\tabletypesize{\scriptsize}
\tablecaption{YSO Classification \label{tbl:ysoclass}}
\tablewidth{0pt}
\tablehead{
\colhead{}    & \colhead{IRAC Colors}                     & \colhead{IRAC \& MIPS Colors}            & \colhead{SED}    \\
\colhead{ID}  & \colhead{[3.6]-[4.5] vs. [5.8]-[8.0]}     & \colhead{[3.6]-[5.8] vs. [8.0]-[24]}     & \colhead{Fitting}  
} 
\startdata
G18-1  & I    & I       & I       \\
G18-2  & I    & I       & I       \\
G18-3  & I    & I       & II      \\
G18-4  & I    & I/HE    & II      \\
G18-5  & I    &  I      & II      \\
G18-6  & I    & \nodata & \nodata \\
G18-7  & II   & \nodata & \nodata \\
G18-8  & II   & \nodata & \nodata \\
G18-9  & II   & II      & III     \\
G18-10 & II   & \nodata & \nodata \\
G18-11 & II   & \nodata & \nodata \\
G18-12 & II   & \nodata & \nodata \\
G18-13 & I/II & \nodata & \nodata \\
G18-14 & I/II & \nodata & \nodata \\
G18-15 & I/II & \nodata & \nodata \\
G18-16 & I/II & \nodata & \nodata \\
G18-17 & I/II & \nodata & \nodata \\
G18-18 & I/II & \nodata & \nodata \\
G18-19 & I/II & \nodata & \nodata \\
G19-1  & I    & \nodata & \nodata \\
G19-2  & I    & \nodata & \nodata \\
G19-3  & I    & I       & I       \\
G19-4  & I    & I       & II      \\
G19-5  & I    & HE      & II      \\
G19-6  & I    & II      & II      \\
G19-7  & I    & I       & I       \\
G19-8  & II   & II      & I       \\
G19-9  & II   & I/II    & II      \\ 
G19-10 & II   & \nodata & \nodata \\
G19-11 & II   & \nodata & \nodata \\
G19-12 & II   & \nodata & \nodata \\
G19-13 & II   & \nodata & \nodata \\
G19-14 & II   & I/II    & II      \\
G19-15 & I/II & \nodata & \nodata \\
G19-16 & I/II & \nodata & \nodata \\
G19-17 & I/II & I/II    & I       \\
\enddata
\tablecomments{HE = Hot Excess \citep{rho06}; I/II = red sources outside of Class I or II areas for IRAC colors or objects in I/II overlap region for IRAC \& MIPS colors; SED fitting done only for sources with [24] data.}
\end{deluxetable}

\clearpage

\begin{deluxetable}{lccccccc}
\tabletypesize{\scriptsize}
\tablecaption{SED Best-Fit YSO Models for IRAC-MIPS Sources \label{tbl:modfit}}
\tablewidth{0pt}
\tablehead{
\colhead{ID} & \colhead{Model \#} & \colhead{Inclination Angle} & \colhead{$A_V$}               & \colhead{Stellar Mass}        & \colhead{Envelope Accretion Rate}                & \colhead{Disk Mass}               & \colhead{YSO Stage} \\
\colhead{}   & \colhead{}         & \colhead{($\degr$)}           & \colhead{(mag.)}                    & \colhead{$M_*$ (M$_\sun$)}    & \colhead{$\dot{M}_e$ (M$_\sun$  yr$^{-1}$)}      & \colhead{$M_d$ (M$_\sun$)}        & \colhead{}
}
\startdata
G18-1  & 3008766 & 18.19 & 14.52 & 3.16 & $8.25\times10^{-5}$ & $7.50\times10^{-2}$ & I   \\
G18-2  & 3009923 & 49.46 & 12.28 & 7.72 & $4.59\times10^{-4}$ & $1.89\times10^{-1}$ & I   \\
G18-3  & 3002489 & 81.37 &  9.75 & 4.15 & 0                   & $6.65\times10^{-4}$ & II  \\
G18-4  & 3012453 & 81.37 & 45.36 & 5.77 & 0                   & $1.03\times10^{-2}$ & II  \\
G18-5  & 3014480 & 81.37 &  8.89 & 4.39 & 0                   & $5.68\times10^{-3}$ & II  \\
G18-9  & 3003037 & 49.46 & 22.31 & 6.90 & 0                   & $2.98\times10^{-8}$ & III \\
G19-3  & 3012136 & 18.19 & 13.20 & 3.41 & $1.27\times10^{-3}$ & $1.41\times10^{-2}$ & I   \\
G19-4  & 3002683 & 81.37 & 12.81 & 4.37 & 0                   & $3.66\times10^{-3}$ & II  \\
G19-5  & 3014289 & 75.52 & 28.43 & 7.65 & 0                   & $1.04\times10^{-1}$ & II  \\
G19-6  & 3018962 & 18.19 &  5.11 & 4.55 & 0                   & $1.39\times10^{-3}$ & II  \\
G19-7  & 3010932 & 18.19 & 14.35 & 7.51 & $4.38\times10^{-4}$ & $7.37\times10^{-3}$ & I   \\
G19-8  & 3008529 & 41.41 &  6.19 & 2.66 & $3.26\times10^{-5}$ & $9.28\times10^{-2}$ & I   \\
G19-9  & 3015777 & 41.41 & 16.34 & 4.93 & $9.86\times10^{-7}$ & $4.52\times10^{-2}$ & II  \\
G19-14 & 3017501 & 75.52 & 16.34 & 8.84 & 0                   & $3.09\times10^{-2}$ & II  \\
G19-17 & 3013269 & 18.19 &  5.34 & 2.14 & $3.13\times10^{-4}$ & $5.09\times10^{-3}$ & I   \\
\enddata
\end{deluxetable}


\begin{thebibliography}{}


\bibitem[Alexander et al.(2012)]{ale12}Alexander, M.J., Kobulnicky,
  H.A., Arvidsson, K., \& Kerton, C.R. 2012, \apj, submitted

\bibitem[Allen et al.(2004)]{all04}Allen, L.E., Calvet, N., D'Alessio,  P. et al. 2004, \apjs, 154, 363

\bibitem[Arvidsson \& Kerton(2011)]{arv11}Arvidsson, K. \& Kerton,  C. R. 2011, \aj, 141, 153

\bibitem[Benjamin et al.(2003)]{ben03} Benjamin R. A., Churchwell, E., Babler, B. L., et al. 2003, PASP, 115, 953

\bibitem[Brogan et al.(2006)]{bro06}Brogan, C.L., Gelfand, J.D., Gaensler, B.M., Kassim, N.E., \& Lazio, T.J.W. 2006, \apjl, 639, 25

\bibitem[Carey et al.(2009)]{car09}Carery, S.J., Noriega-Crespo, A., Mizuno, D.R., et al. 2009, PASP, 121, 76

\bibitem[Churchwell(1975)]{chu75}Churchwell, E. 1975, in Lecture Notes in Physics 42, HII Regions and Related Topics, ed. T.L. Wilson \& D. Downes, (New York: Springer-Verlag), page

\bibitem[Churchwell et al.(2006)]{chu06}Churchwell, E., Povich, M.S., Allen, D. et al. 2006, \apj, 649, 759

\bibitem[Churchwell et al.(2009)]{chu09}Churchwell, E., Babler, B.L.,  Meade, M.R., et al. 2009, \pasp, 121, 213 

\bibitem[Clemens(1985)]{cle85}Clemens, D.P. 1985, \apj, 295, 422

\bibitem[Conti \& Crowther(2004)]{con04}Conti, P.S. \& Crowther,  P.A. 2004, \mnras, 355, 899

\bibitem[Crowther(2005)]{cro05}Crowther, P.A. 2005, in IAU Symp. 227,  Massive Star Birth:  A Crossroads of Astrophysics, ed. R. Cesaroni,  M. Felli, E. Churchwell, \& C.M. Walmsley, (Cambridge: Cambridge University Press), 389.

\bibitem[Emerson(1988)]{eme88}Emerson, J.P. 1988, in NATO ASIC  Proc. 241, Formation and Evolution of Low Mass Stars,  ed. A.K. Dupree \& M.T.V.T. Lago (Dordrecht: Kluwer), 193

\bibitem[Engelbracht et al.(2007)]{eng07}Engelbracht, C.W., Blaylock,  M., Su, K.Y.L., et al. 2007, \pasp, 119, 994

\bibitem[Finkenzeller \& Mundt(1984)]{fin84}Finkenzeller, U. \& Mundt,  R. 1984, \aaps, 55, 109

\bibitem[Giard et al.(1994)]{gia94}Giard, M., Bernard, J.P.,  Lacombe, F., Normand, P., \& Rouan, D. 1994, \aap, 289, 524

\bibitem[Hartmann et al.(2005)]{har05}Hartmann, L., Megeath, S.T., Allen, L. et al. 2005, \apj, 629, 881

\bibitem[Helfand et al.(2006)]{hel06}Helfand, D. J., Becker, R. H., White, R. L., Fallon, A., \& Tuttle, S. 2006, \aj, 131, 2525
 
\bibitem[Higgs et al.(1997)]{hig97}Higgs, L.A., Hoffmann, A.P., \&  Willis, A.G. 1997, in ASP Conf. Ser. 125, Astronomical Data Analsis Software and Systems  VI, ed. G. Hunt \& H.E. Payne (San Francisco, CA: ASP), 58.

\bibitem[Jackson et al.(2006)]{jac06} Jackson, J. M., Rathborne, J. M., Shah, R. Y., et al. 2006, \apjs, 163, 145

\bibitem[Johanson \& Kerton(2009)]{joh09}Johanson, A.K. \& Kerton, C.R. 2009, \aj, 138, 1615

\bibitem[Kerton et al.(1999)]{ker99}Kerton, C.R., Ballantyne, D.R., \&  Martin, P.G. 1999, \aj, 117, 2485

\bibitem[Kerton et al.(2008)]{ker08}Kerton, C.R., Arvidsson, K., Knee, L.B.G., \& Brunt, C. 2008, \mnras, 385, 995

\bibitem[Kauffmann et al.(2008)]{kau08}Kauffmann, J., Bertoldi, F.,  Bourke, T.L., Evans, N.J., \& Lee, C.W. 2008, \aap, 487, 993 

\bibitem[Lockman(1989)]{loc89} Lockman, F. J. 1989, \apjs, 71, 469

\bibitem[Lucas et al.(2008)]{luc08}Lucas, P.W., Hoare, M.G., Longmore, A., et al. 2008, \mnras, 391, 136

\bibitem[Matsakis et al.(1976)]{mat76}Matsakis, D.N., Evans, N.J.,  Sato, T., \& Zuckerman, B. 1976, \aj, 81, 172

\bibitem[Megeath et al.(2004)]{meg04}Megeath, S.T., Allen, L.E.,  Gutermuth, R.A., et al. 2004, \apjs, 154, 367

\bibitem[Mezger \& Henderson(1967)]{mez67}Mezger, P.G. \& Henderson,  A.P. 1967, \apj, 147, 471 

\bibitem[Osterbrock(1989)]{ost89}Osterbrock, D.E. 1989, Astrophysics of Gaseous Nebulae and Active Galactic Nuclei (Sausalito, CA:  University Science Books)

\bibitem[Pohl et al.(2008)]{poh08}Pohl, M., Englmaier, P., \&  Bissantz, N. 2008, \apj, 677, 283

\bibitem[Rho et al.(2006)]{rho06}Rho, J., Reach, W.T., Lefloch, B., \&  Fazio, G.G. 2006, \apj, 643, 965

\bibitem[Robitaille et al.(2007)]{rob07}Robitaille, T.P., Whitney,  B.A., Indebetouw, R., \& Wood, K. 2007, \apjs, 169, 328

\bibitem[Rosolowsky et al.(2010)]{ros10} Rosolowsky, E., Dunham,  M. K., Ginsburg, A., et al. 2010, \apjs, 188, 123

\bibitem[Rubin(1968)]{rub68}Rubin, R. H. 1968, \apj, 154, 391

\bibitem[Russeil(2003)]{rus03}Russeil, D. 2003, \aap, 397, 133

\bibitem[Schaerer \& de Koter(1997)]{sch97}Schaerer, D. \& de Koter,  A. 1997, \aap, 322, 598

\bibitem[Simpson et al.(2012)]{sim12}Simpson, R. J., Povich, M. S.,  Kendrew, S., et al. 2012, \mnras, 424, 2442

\bibitem[Skrutskie et al.(2006)]{skr06}Skrutskie, M.F., Cutri, R.M., Stiening, R., et al. 2006, \aj, 131, 1163

\bibitem[Spitzer(1978)]{spi78}Spitzer, L. 1978, Physical Processes in the Interstellar Medium (New York: John Wiley \& Sons)

\bibitem[Stil et al.(2006)]{sti06} Stil, J. M., Taylor, A. R., Dickey, J. M., et al. 2006, \aj, 132, 1158

\bibitem[The et al.(1994)]{the94}The, P.S., de Winter, D., \& Perez,  M.R. 1994, \aaps, 104, 315

\bibitem[Tielens et al.(1993)]{tie93}Tielens, A.G.G.M., Meixner, M.M.,  van der Werf, P.P., et al. 1993, Science, 262, 86

\bibitem[Ward-Thompson(2002)]{war02}Ward-Thompson, D. 2002, Science,
  295, 76

\bibitem[Westerhout(1958)]{wes58} Westerhout, G. 1958, BAN, 14, 215


\end{thebibliography}
\end{document}